\documentclass[aps,amssymb,amsmath,superscriptaddress,pra,twocolumn,showpacs]{revtex4-1}
\usepackage{graphicx}
\usepackage{subfigure}
\usepackage{amsfonts}
\usepackage{extarrows}
\begin{document}
\title{Quantum phase transitions in spin-1 XXZ chains with rhombic single-ion anisotropy}
\author{Jie Ren}
\email{jren@cslg.edu.cn}
\affiliation{Department of Physics and Jiangsu Laboratory of Advanced
Functional Material, Changshu Institute of Technology, Changshu 215500, China}

\author{Yimin Wang}
\affiliation{College of Communications Engineering, The Army Engineering University of PLA, Nanjing 210007, China}

\author{Wen-Long You}
\email{wlyou@suda.edu.cn}
\affiliation{College of Physics, Optoelectronics and Energy, Soochow University, Suzhou, Jiangsu 215006, China}

\date{\today}
\begin{abstract}
We explore numerically the inverse participation ratios in the ground-state of one-dimensional spin-1 XXZ chains with the rhombic single-ion anisotropy.  By employing the techniques of density matrix renormalization group, effects of the rhombic single-ion anisotropy on various information theoretical measures are investigated, such as the fidelity susceptibility, the quantum coherence and the entanglement entropy. Their relations with the quantum phase transitions are also analyzed. The phase transitions from the Y-N\'{e}el phase to the large-$E_x$ or the Haldane phase can be well characterized by the fidelity susceptibility. The second-order derivative of the ground-state energy indicates all the transitions are of second order. We also find that the quantum coherence, the entanglement entropy, the Schmidt gap and the inverse participation ratios can be used to detect the critical points of quantum phase transitions. Results drawn from these quantum information observables agree well with each other. Finally we provide a ground-state phase diagram as functions of the exchange anisotropy $\Delta$ and the rhombic single-ion anisotropy $E$.
\end{abstract}
\maketitle

\section{introduction}
\label{sec:intorduction}
Quantum phase transition (QPT) is a very important phenomenon in
condensed-matter physics, and it happens at zero temperature by
tuning one or more external parameters in the system's Hamiltonian \cite{Sachdev}. Among them, spin $S=1$ antiferromagnetic Heisenberg chain has been extensively studied both experimentally and theoretically \cite{Nijs,Chen,Boschi,Botet,Nomura,Sakai,Darriet,Buyers,Steiner}. It is noted that the ground state is in Haldane phase, which has nonlocal string order. It is gapped between a spin-singlet ground state and a spin-triplet excited state, and has gapless entanglement spectra. These properties can be used to characterize the Haldane phase.
However, an ideal one-dimensional (1D) spin-1 system is accompanied by the interchain interactions and magnetic anisotropy, which may partially or completely suppress the excitation gap and thus lead to detection of an observation of long-range order in a quantum disordered magnet. The rhombic single-ion anisotropy was discovered in some materials, such as Y$_2$BaNiO$_5$ \cite{Darriet},  NBYC \cite{Batchelor} and heterobimetallic complexes \cite{Lillo09}. On the other hand, since the adventure of quantum engineering has made rapid progress in recent years,
the exchange interaction and the magnetic anisotropy can be modulated through Kondo physics \cite{Jacobson,Liu01,Otte}, scanning tunneling
microscope \cite{Khajetoorians,Hirjibehedin,Loth,Khajetoorians01}, and the exchange-biased quantum tunneling \cite{Wernsdorfer,Nguyen}. However, the effect of rhombic single-ion anisotropy lacks a complete theoretical understanding.


The competitions among various physical mechanisms will induce QPTs and thus essentially enrich the ground-state phase diagram of the
spin model. To characterize various phases, which go beyond the Haldane phase of spin-1 Heisenberg Hamiltonian, we adopt multiple theoretical measures to identify the critical points and quantum phases. Recently, various exogenous approaches inherited from quantum information have been exploited to measure the curvature of the many-body ground states.
Much effort has been put into the study of quantum critical
phenomena in spin chains in terms of quantum information theory. Two well-known and widely-studied measures of quantum correlations are quantum coherence \cite{Baumgratz} and entanglement entropy (EE) \cite{Amico}. Another concept was frequently referred to 
as fidelity susceptibility (FS), which measures the changing rate between two closest states. FS diverges at
the critical points in the thermodynamic limit \cite{Gu}. The ground-state quantum correlations and FS were deemed to be capable of qualifying QPTs in strongly correlated systems \cite{Abasto,Buonsante,You,You2,Zhou01,legeza,Ren01,Ren02,You12,Liu12,Ren03,Ren05,Ren06,Malvezzi}, since QPTs are intuitively associated with an abrupt change in the structure of the ground-state wave function. This primary observation motivates researchers to use quantum coherence, EE and FS to predict QPTs. The scaling relation of FS 
was proposed for the spin-1 XXZ spin chain with a single-site anisotropy term \cite{Tzeng}. Through a proper finite-size scaling analysis, the results from both FS and EE agree with the findings in the previous results \cite{Tzeng1}. The effect of rhombic single-ion anisotropy in the $S=1$ Haldane chain was lately investigated and a precise ground-state phase diagram was identified \cite{Tzeng3}. However, the common XXZ anisotropy was not taken into account. Such exchange anisotropy induces a strong dependence of the magnetization process on the magnetic field direction between the in-plane (XY) and out-of-plane (Ising) exchange interactions in spin. To this end, it would be interesting to discuss the effect of rhombic single-ion anisotropy in the $S=1$ XXZ chain.

In this paper, we make use of the ground-state EE and FS, as well as nonlocal correlations, to analyze the QPTs in the 1D spin-1 XXZ chains with rhombic single-ion anisotropy. The Hamiltonian and the details of numerical methods as well as the measurements
are shown in Sec. \ref{sec:Hamiltonian}. In Sec. \ref{sec:results}, results of all theoretical measures, including quantum correlation measures and the FS as well are presented. A discussion is provided in the last section.

\begin{figure*}[t]
\centering
\includegraphics[width=1.00\textwidth]{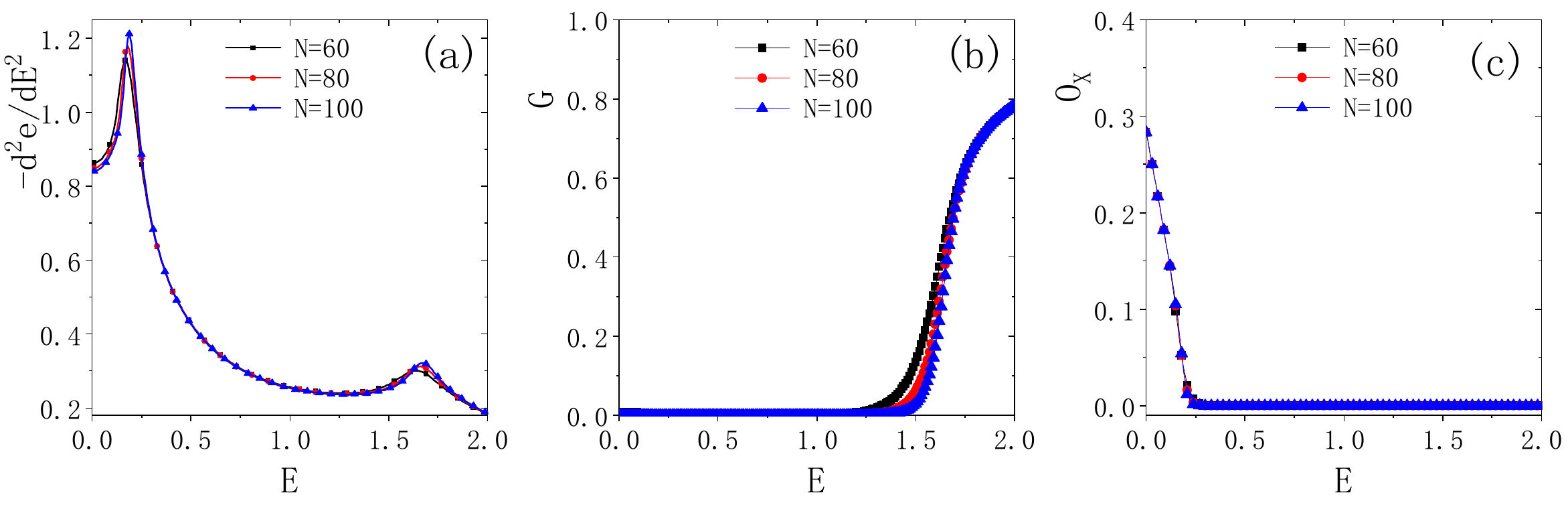}
\caption{\label{fig1}(a) Second-order derivative of the ground-state energy density, (b) the Schmidt gap $G$ and (c) the string order parameter are plotted as a function of the rhombic single-ion anisotropy $E$ for different system sizes with $\Delta=1$.
\label{fig1}}
\end{figure*}

\section{Hamiltonian and Measurements}
\label{sec:Hamiltonian}
The Hamiltonian of a 1D spin-1 XXZ chain with the rhombic single-ion anisotropy is given by
\begin{eqnarray}
\label{eq1}
&H=\displaystyle{  \sum_{i=1}^{N}}J(S_i^x S_{i+1}^x+S_i^y S_{i+1}^y+\Delta S_i^z S_{i+1}^z)\nonumber \\  &+ \displaystyle{ \sum_{i=1}^{N}}E[(S_i^x)^2-(S_i^y)^2], \label{Hamiltonian}
\end{eqnarray}
where $S_i^\alpha(\alpha=x,y,z)$ are spin-$1$ operators on the $i$-th site and $N$
is the length of the spin chain. The parameter $J$ denotes the antiferromagnetic coupling, and $J$=1 is assumed hereafter in the paper. The parameters $\Delta$ and $E$ are the exchange anisotropy and the rhombic single-ion anisotropy, respectively. The open boundary condition is assumed in the paper.

In the following we make use of  the density matrix
renormalization group (DMRG)~\cite{white,U01,U02} method, with which the
ground state of the 1D system in large sizes can be calculated with
very high accuracy.  More precisely, we implement GPU speeding up to Matlab
code for the finite-size DMRG with double precision data and four sweeps.
The maximum number of eigenstates kept is
$m=200$ during the procedure of basis truncation, and such truncation guarantees the converging error is smaller than $10^{-8}$ for system sizes up to $N=200$. With such accurate performance calculation, we can precisely
analyze the QPTs through various theoretic measures.

In the absence of the rhombic single-ion anisotropy, the integrable spin-1 Heisenberg supports a gapped ground state and
serves as a useful example for Haldane conjecture \cite{Haldane} and other concepts such as a hidden $Z_2$$\times$$Z_2$
symmetry breaking and symmetry protected topological order. The nonlocal  string  order  captures  the  hidden  symmetry  breaking  in  the  Haldane phase  of  the  1D spin-1  Heisenberg model  and can be characterized by the string order parameter (SOP), whose definition
is given by \cite{Nijs}
\begin{equation}
O_{x} = - \lim_{(j-i) \to \infty} [S_i^{x} \exp(i \pi \sum_{i<l<j} S_l^{x}) S_j^{x}].
\label{eq2}
\end{equation}
The SOP characterizes the topological order only within the Haldane phase, as the measurement $O_{x}$ is nonzero in the Haldane phase and zero elsewhere. To explore the effects of
the exchange anisotropy and the rhombic single-ion anisotropy, we consider another quantity of interest, i.e., the inverse participation ratio (IPR).
The IPR entirely depends on the choice of basis. In a specific $D$-dimensional bases $|\varphi_k\rangle$ ($k=1,2,\cdots,D$), the IPR of the state $|\psi_0\rangle$ is defined as
\begin{equation}
T=\left[\frac{\sum_{k=1}^D|c_k|^4}{\sum_{k=1}^D|c_k|^2}\right]^{-1}.
\label{eq3}
\end{equation}
Here $|\psi_0\rangle=\sum_{k=1}^D c_k|\varphi_k\rangle$. 
In the following we focus on the ground state of Eq.(\ref{Hamiltonian}).
The IPR of the ground state reaches a minimal value $T_{min}=1$ when the ground-state coincides exactly a single basis state, and attains a maximal value $I_{max}=D$ when the ground state is uniform in the selective bases. The IPRs can quantify the extent of distribution over the preferential bases. For a set of one-particle states in real, a large IPR is associated with a delocalized state, whereas a small one is relayed to a localized state \cite{Izrailev}. 

As the external parameter (e.g., $E$ or $\Delta$) varies across a critical point, the ground-state wave function undergoes a sudden change in the wake of QPT, accompanied by a rapid alteration in the quantum correlation. Quantum coherence is a resurgent concept in quantum theory and acts as a manifestation of the quantum superposition principle. The Wigner-Yanase skew information (WYSI), which we adopt as a measure
of coherence, is given by \cite{Wigner}
\begin{equation}
I(\rho,K)=-\frac{1}{2}\textrm{Tr}([\sqrt{\rho},K]^2),
\label{eq4}
\end{equation}
where the density matrix $\rho$ describes a quantum state, $K$ plays a role of
an observable, and [.,.] denotes the commutator. It can be noted that as the skew information reduces
to the variance $V(\rho,K)=\textrm{Tr} \rho K^2-(\textrm{Tr} \rho K)^2$ for pure states, and it is upper bounded by the variance for mixed states.
We will simply refer to $I(\rho ,K)$ as $K$ coherence in the paper.

For a system composed of $A$ and $B$ subsystems, the bipartite EE can be chosen as an alternative measurement of the quantum correlation,
\begin{equation}
\label{eq5}S_L=-\textrm{Tr}(\rho_{A}\log_2\rho_{A}),
\end{equation}
in which the reduced density matrix of subsystem $A$ part is $\rho_{A}=\textrm{Tr}_{B}(|\psi_0\rangle \langle \psi_0|)$
for the ground state $|\psi_0\rangle$. 
One convenient choice of the subsystem $A$ is composed of the first $L$ sites, and the subsystem $B$
is the rest of the system.  In addition to EE, the Schmidt gap can also be used to describe the QPTs \cite{Chiara}. It is defined as
\begin{equation}
\label{eq6}G=g_1-g_2,
\end{equation}
where $g_1$ and $g_2$ are the first and the second largest eigenvalues of the reduced density matrix $\rho_{A}$, respectively. The
Schmidt gap has been shown to act as an order parameter and capture quantum phase transitions in the ground
state of zero temperature systems \cite{Chiara,Bayat}.


On the other hand, fidelity susceptibility measures the changing rate of similarity between the two closest states as the external parameter $\lambda$ is tuned, which is defined as\cite{You}:
\begin{equation}
\label{eq7}
|\langle
\psi_0(\lambda)|\psi_0(\lambda+ \delta \lambda)\rangle|  =1-\frac{\chi(\lambda )}{2}\delta \lambda^2 + O(\delta \lambda^3),
\end{equation}
where $\delta \lambda $ is an infinitesimal distance. The fidelity susceptibility is an information metric in the $d$-dimensional parameter, which has a gravity dual with the spatial volume of the
Einstein-Rosen bridge in anti-de Sitter (AdS) \cite{Miyaji15}.
The divergence of  $\chi(\lambda )$ can directly locate the critical points. The efficiency in identifying the continuous QPTs is exactly convincing \cite{Cozzini,You,You2}.
\begin{figure*}[t]
\includegraphics[width=0.65\textwidth]{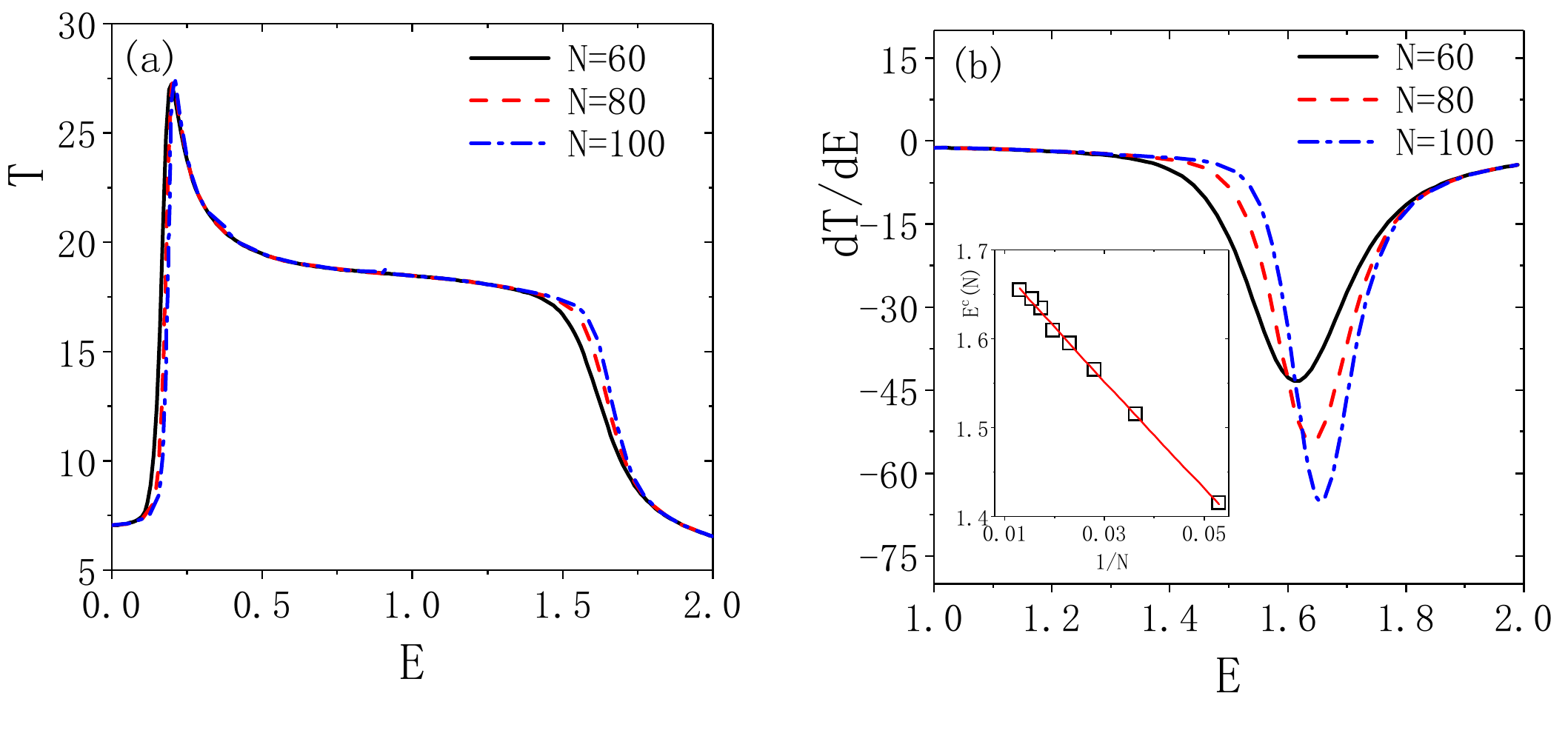}
\caption{(a)Inverse participation ratio $T$ of the ground state and (b) its first-order derivative are plotted as a function of the rhombic single-ion anisotropy $E$ for different system sizes $N$ with $\Delta=1.0$. Inset in (b) shows finite-size scaling of $E_{c}$ of the derivative of the inverse participation ratios as a function of $N^{-1}$. The line is the numerical fitting.
\label{fig2}}
\end{figure*}

\begin{figure}[t]
\includegraphics[width=0.33\textwidth]{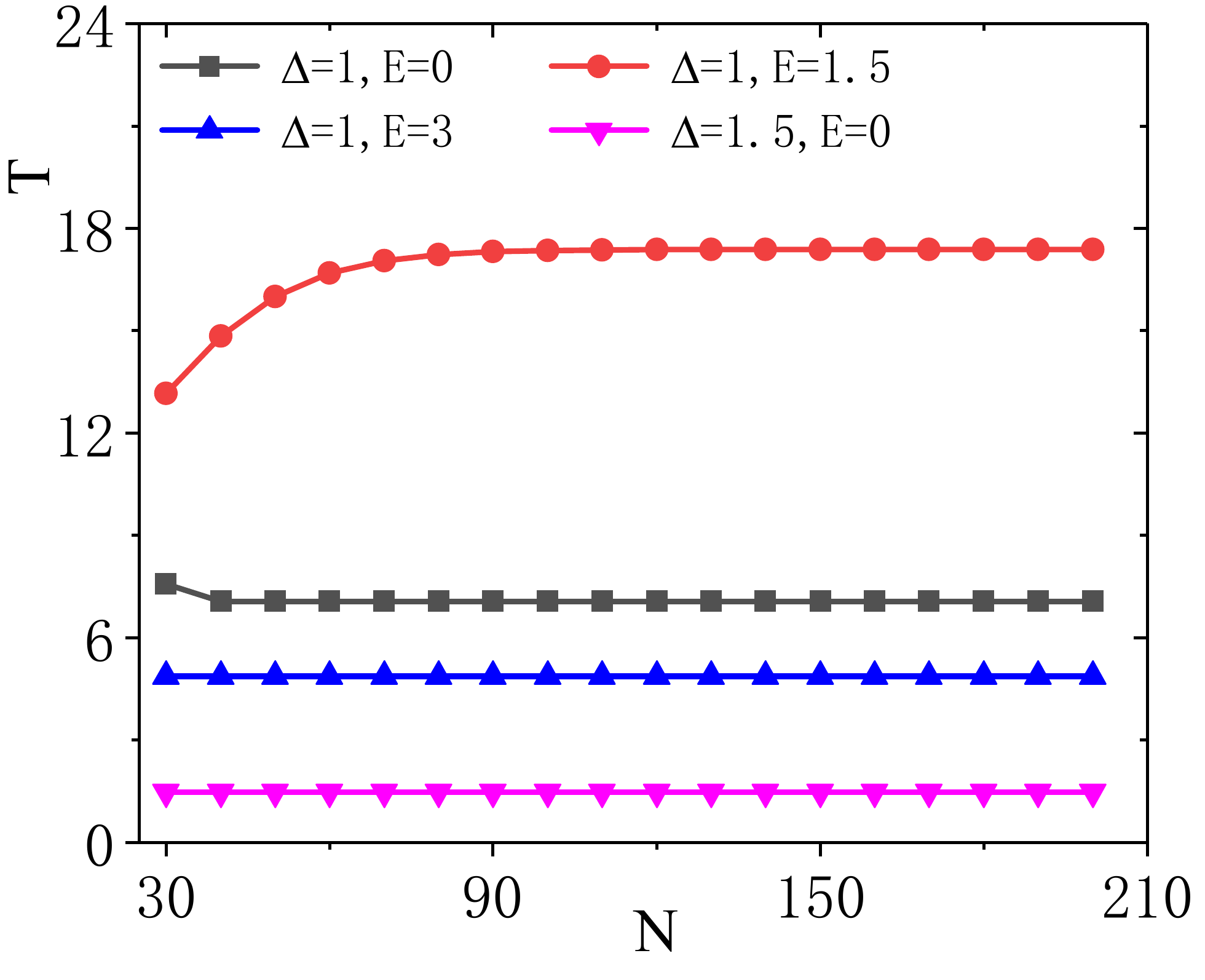}
\caption{Inverse participation ratio $T$ is plotted as a function of system size $N$. A set of selected parameters are located in one of four phases, respectively.
\label{fig2_scaling}}
\end{figure}

\begin{figure}[t]
\includegraphics[width=0.525\textwidth]{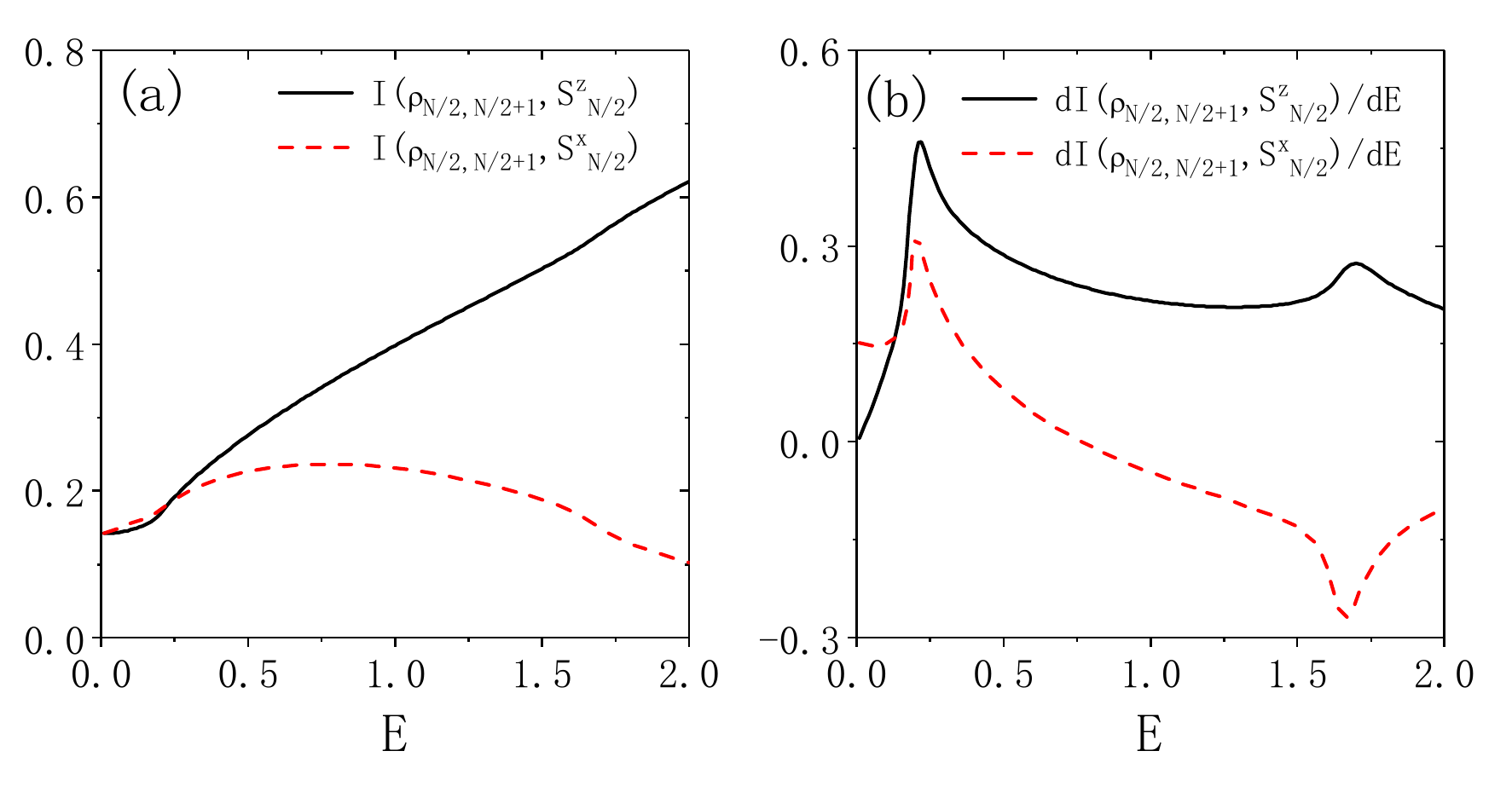}
\caption{(a) Two-site QC $I(\rho_{N/2,N/2+1},S_{N/2}^x)$, $I(\rho_{N/2,N/2+1},S_{N/2}^z)$ and (b) their first-order derivatives are plotted as a function of the rhombic single-ion anisotropy $E$ for $N=100$.
\label{fig3}}
\end{figure}

\begin{figure}
\includegraphics[width=0.55\textwidth]{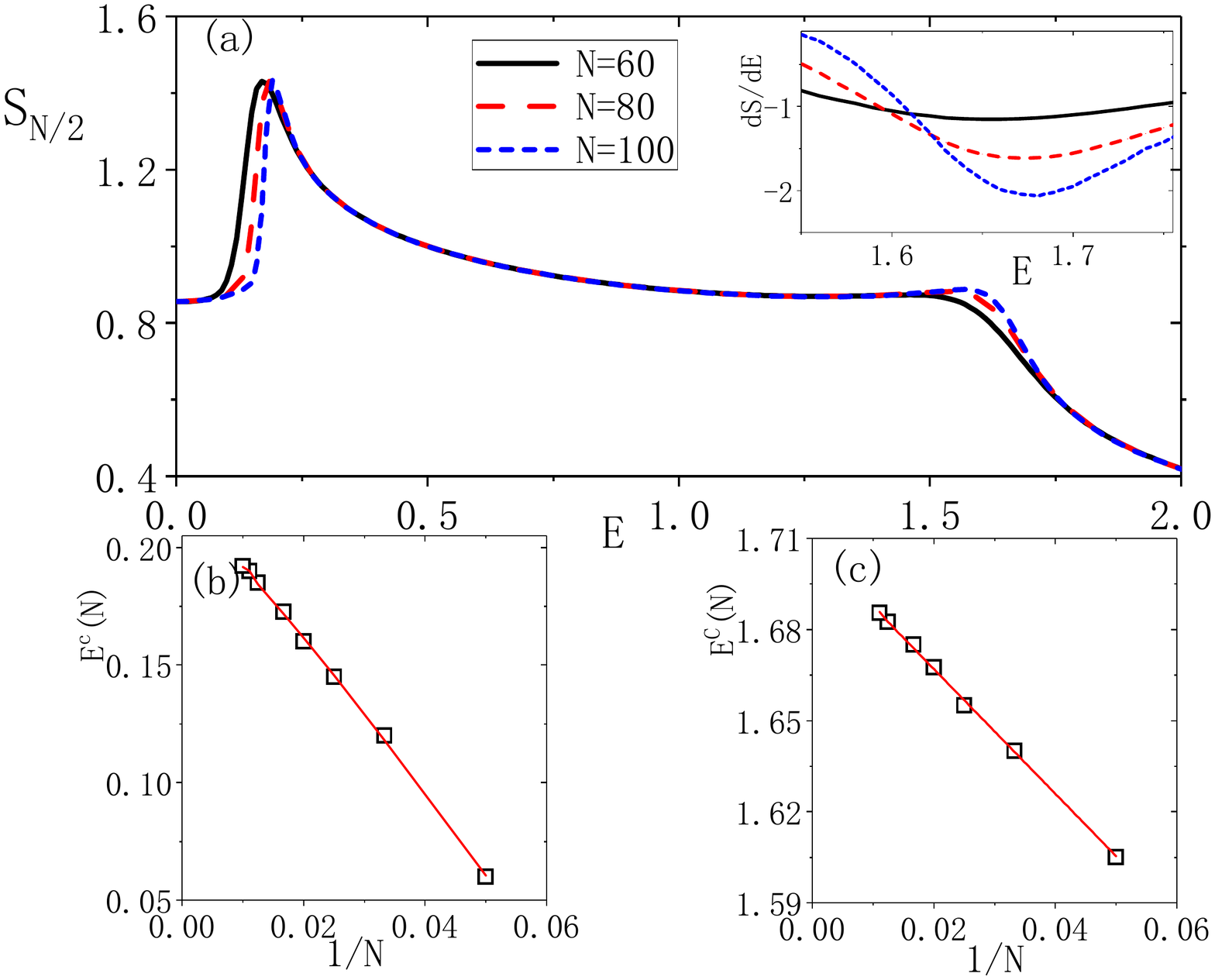}
\caption{ (a) Entanglement entropy is plotted as a function of the rhombic single-ion anisotropy $E$ for different system sizes $N$ with $\Delta=1$. Inset: the first-order derivative of the entanglement entropy. (b) Finite-size scaling of $E^{c}$  of the first extreme point. (c) Finite-size scaling of $E^{c}$ obtained from the first-order derivative of entanglement entropy at the second extreme point. The lines are the fitted lines.
\label{fig4} }
\end{figure}

\begin{figure}[t]
\includegraphics[width=0.450\textwidth]{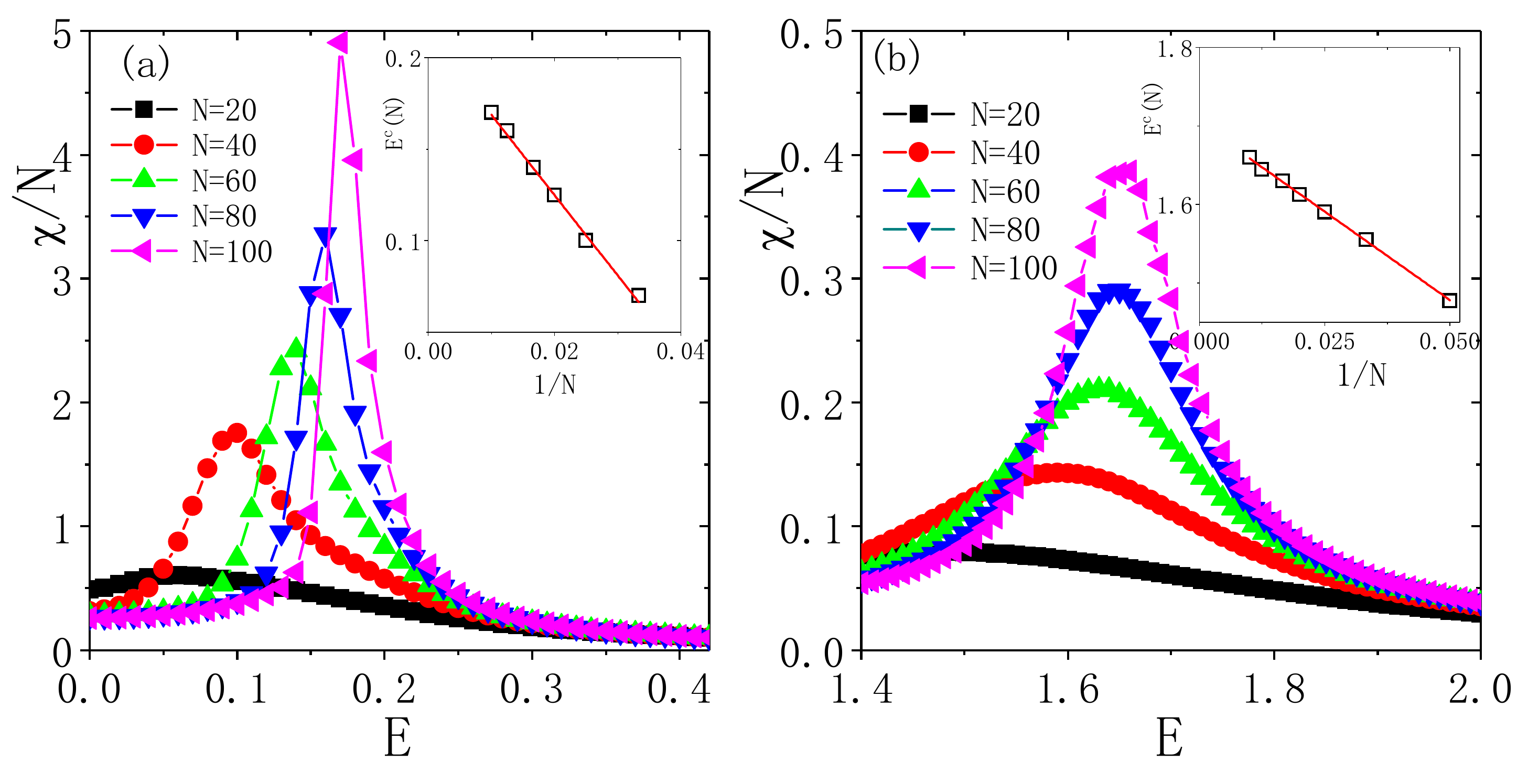}
\caption{Fidelity susceptibility per site is plotted as a function of the rhombic single-ion anisotropy $E$ for different system sizes $N$ with $\Delta=1$. Insets show the finite-size scaling of $E^{c}$ in terms of the fidelity susceptibility versus $N^{-1}$. The lines are the numerical fittings.
\label{fig5}}
\end{figure}

\section{Numerical results}
\label{sec:results}

\subsection{Case of $\Delta=1$}

By means of DMRG, we study the ground-state energy and other relevant quantities. The second-order derivative of the ground-state energy density is shown in Fig. \ref{fig1}(a). When $E$ is small, the system is in the Haldane phase, as is diagnosed by the nonzero SOP in Fig. \ref{fig1}(c).  As $E$ increases, the SOP becomes vanishing and the second-order derivative of the ground-state energy diverges at $E_{1}^{c}=0.214$, implying a QPT occurs.
The system enters into a Y-N\'{e}el phase. After $E$ exceeds $E_{2}^{c}=1.717$\cite{Tzeng3}, the system moves into a large-$E_x$ phase.
The divergences at criticalities imply that both transitions are of second order. Furthermore, the Schmidt gap of the reduced density matrix by cutting a $N$-site chain into two halves $(L = N/2)$ are demonstrated. The Schmidt gap labeled by $G$ is plotted as a function of $E$ for different system sizes in Fig. \ref{fig1}(b). The Schmidt gap quantifies the competition between the two dominant states on either side of the partition. It is found that the Schmidt gap is gapped in large-$E_x$ phase and it is gapless in both the Haldane phase and the Y-N\'{e}el phase. So it can not distinguish the transition between the Y-N\'{e}el phase and the Haldane phase, but can sense the transition between the Y-N\'{e}el phase and large-$E_x$ phase.

In Fig. \ref{fig2}(a), we plot the IPRs labeled by $T$ of the ground state. When the system is in the Haldane phase, the IPRs are small, almost size-independent; see Fig. \ref{fig2_scaling}.
 The IPRs grow sharply with increasing $E$ up to a maximum. When $E$ further goes up, the IPRs drop and the system turns to the Y-N\'{e}el phase \cite{Tzeng3}. One can find the maximum of the peak is size-independent. We identify this peak as the indication of the critical point $E_{1}^{c}=0.21$. The further increase of $E$ pushes the system into the large-$E_x$ phase, and one can find that the IPRs experience a sudden change when $E$ crosses a critical value, implying that the first-order derivative can capture the quantum critical points. This can be understood by through
 \begin{eqnarray}
\label{eq8}
\frac{\partial^2 e}{\partial E^2}=\frac{1}{N}[\mathrm{Tr}(\sum_{i,j}\frac{\partial^2 H_{i,j}}{\partial E^2}\rho)+\mathrm{Tr}(\sum_{i,j}\frac{\partial H_{i,j}}{\partial E}\frac{\partial \rho}{\partial E})].
 \end{eqnarray}
The discontinuity in the second-order derivatives of the energy density $e$ requires the divergence of at least one of the derivatives
$\frac{\partial \rho}{\partial E}$ at the critical points \cite{wu}. The first derivative of IPRs are plotted as a function of the rhombic single-ion anisotropy $E$ for different system sizes with $\Delta=1.0$ in Fig. \ref{fig2}(b). The pesudocritical points on finite-size system $E^c(N)$ are below the true critical points $E^c(\infty)$, which can be extrapolated by a power-law scaling:
\begin{equation}
E^c(N) \sim E^c(\infty)+aN^b.
\label{eq9}
\end{equation}
We fit the locations of the extremes by Eq. (\ref{eq9}), which are shown in the inset of Fig. \ref{fig2}(b). We find that $E_{2}^{c}=1.723, a_1=-5.413, b_1=-0.956$. Moreover, when the system is in the Y-N\'{e}el phase and the large-$E_x$ phase, the IPRs also saturate with respect to the system size $N$, see Fig. \ref{fig2_scaling}. There is negligible difference between $N=150$ and $N=200$. This result is in stark contrast to the participation ratios summing over all the eigenstates in Ref. [\onlinecite{Misguich}].

The WYSI can be used as an efficient measure to quantify quantum coherence (QC), but it is rarely adopted to detect QPTs.
Here we plot the central two-site QC $I(\rho_{N/2,N/2+1},S_{N/2}^x)$, $I(\rho_{N/2,N/2+1},S_{N/2}^z)$ as a function of the rhombic single-ion anisotropy $E$ for $N=100$ in Fig. \ref{fig3}(a). It is found that $I(\rho_{N/2,N/2+1},S_{N/2}^z)$ increases with $E$ and $I(\rho_{N/2,N/2+1},S_{N/2}^x)$ reaches a maximal value at $E=0.79$. According to the argument proposed in Refs. \cite{Osterloh,Popp}, the first-order derivatives of the local QC are capable of detecting different types of QPTs in many-body systems \cite{Lin}. The extreme points shown in Fig. \ref{fig3}(b) mark the corresponding critical points, although there is a small difference between $I(\rho_{N/2,N/2+1},S_{N/2}^x)$ and $I(\rho_{N/2,N/2+1},S_{N/2}^z)$.
We also investigate the EE between the rightmost half part and the rest. The entanglement is plotted as a function of the rhombic single-ion anisotropy $E$ for different system sizes in Fig. \ref{fig4}(a). The EE initially grows up as $E$ increases, and then undergoes a overturn when the system transits from the Haldane phase to the Y-N\'{e}el phase. The EE decreases gradually with increasing $E$.  A further increase of $E$ induces the system into the Large-$E_x$ and the EE declines more dramatically with respect to the increase of $E$. One finds that the EE shows remarkable finite-size effects in the Haldane phase and around the critical points.
We fit the location of the first peak by Eq. (\ref{eq9}). We plot the location of the maximum EE as a function of $1/N$ and show the numerical fit in Fig. \ref{fig4}(b). We obtain that $E_{1}^{c}=0.217$. 
This was confirmed by previous results \cite{Tzeng3}.
As for the second critical point, the criticality can be seized by the first-order derivative of the EE, which is similar to the previous results of IPRs. To this end, the quantum critical points sometimes are determined by the positions at which the first-order derivative of the EE takes extreme values \cite{Ren03,Ren01,Liu}. The first-order derivative of entanglement is plotted as a function of $E$ for different system sizes in the inset of Fig. \ref{fig4}(a). The valley in the first derivative of the EE can detect the critical points. The positions of the valley can be extrapolated to the thermodynamic limit, which is shown in Fig. \ref{fig4}(c). The critical point $E^c_2=1.712$ is found.

By means of DMRG, we calculate the ground-state FS with various system sizes $N$ up to $100$. The ground-state FS per site $\chi/N$ is plotted as a function of the rhombic single-ion anisotropy $E$ for different sizes in Fig. \ref{fig5}.
Two peaks of the FS are observed for positive rhombic single-ion anisotropy. The peak value of $\chi/N$ increases when the system size rises.  The location of peaks moves to a slightly higher $E$ up to a particular value as the system size $N$ increases.
Here, the scaling of those extreme points of the FS 
is also investigated. We find that the positions of the maximal points can also be fitted by a formula $E \sim E^c+a/N$, where $a$ is a constant.
The results for the locations of the FS can be used to predict the QPT points in the thermodynamic limit. In the inset of Fig. \ref{fig5},
we plot the location of the maximum FS as a function of $1/N$ and draw the numerical fit in red. We obtain $E_{1}^{c}=0.212, a_2=-4.38$ and  $E_{2}^{c}=1.705,a_3= -4.50$.

\begin{figure}[t]
\includegraphics[width=0.350\textwidth]{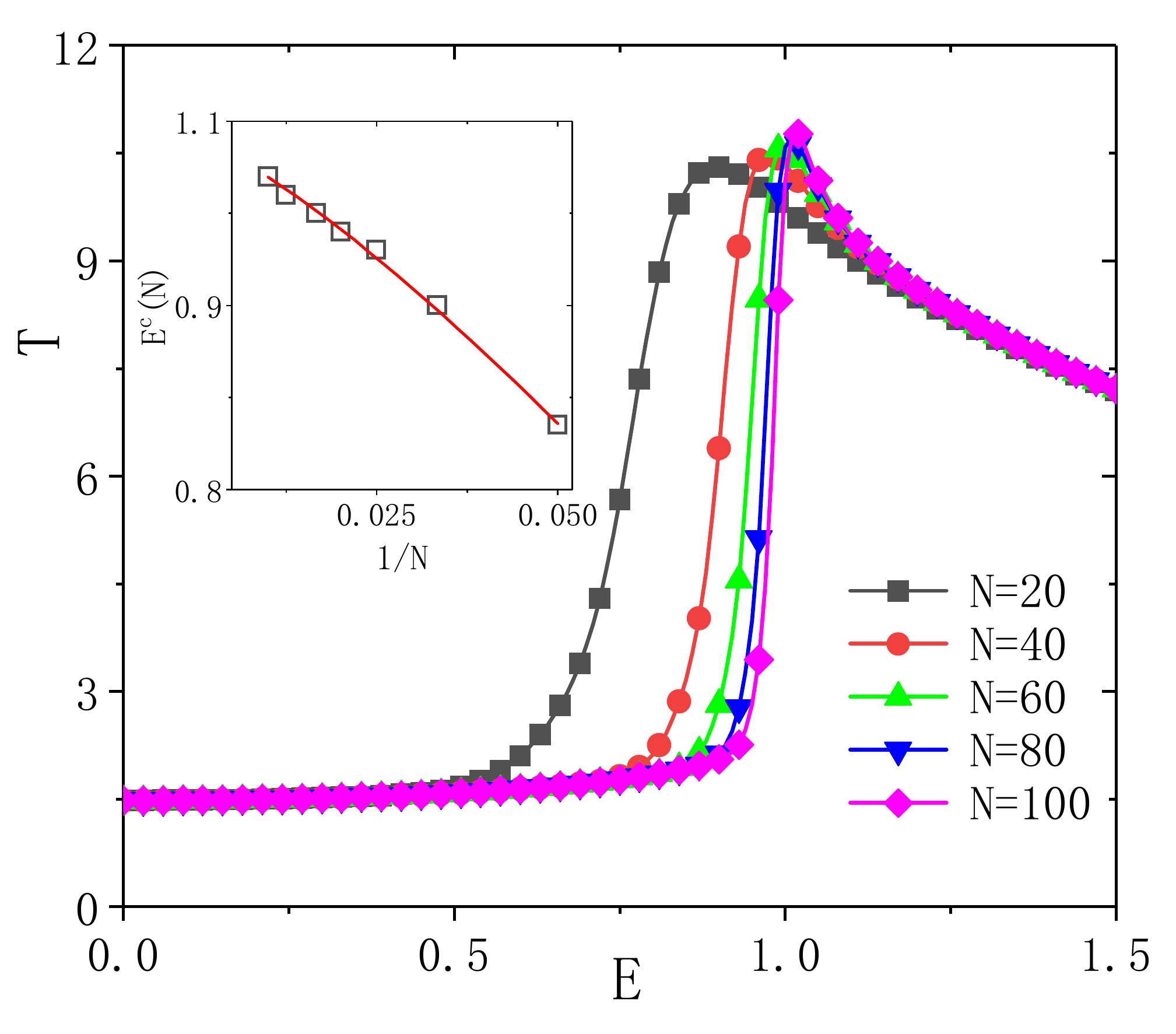}
\caption{ Inverse participation ratio $T$ of the ground state is plotted as a function of the rhombic single-ion anisotropy $E$ for different system sizes with $\Delta=1.5$. Inset: Finite-size scaling of $E^{c}$ of the extreme point of the inverse participation ratios.
\label{fig6}}
\end{figure}

\begin{figure}[t]
\includegraphics[width=0.53\textwidth]{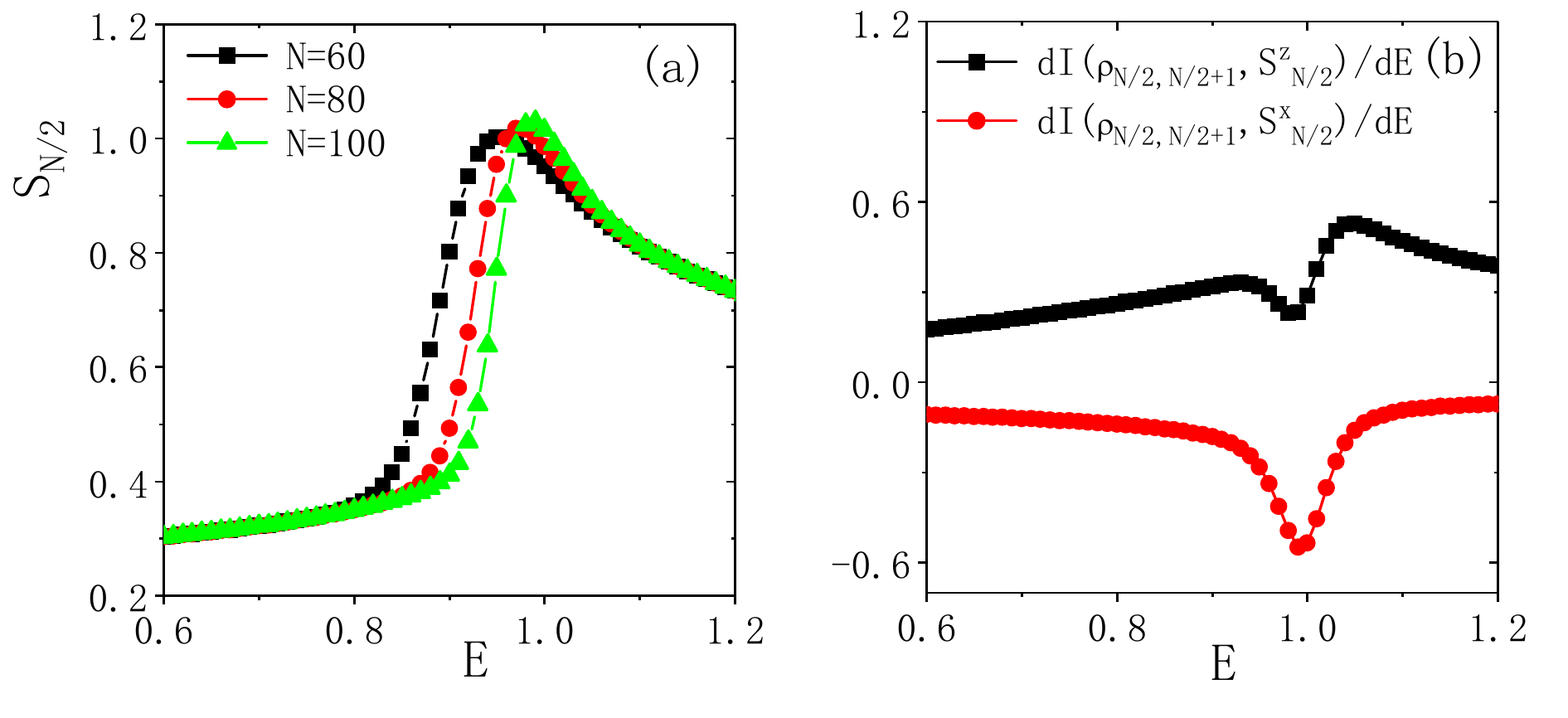}
\caption{(a) Entanglement entropy is plotted as a function of the rhombic single-ion anisotropy $E$ for different system sizes with $\Delta=1.5$. (b) The first-order derivative of the two-site QC $I_{N/2,N/2+1}$ is plotted as a function of $E$ with $\Delta=1.5$ for $N=100$.
\label{fig7}}
\end{figure}

\begin{figure*}[t]
\includegraphics[width=0.75\textwidth]{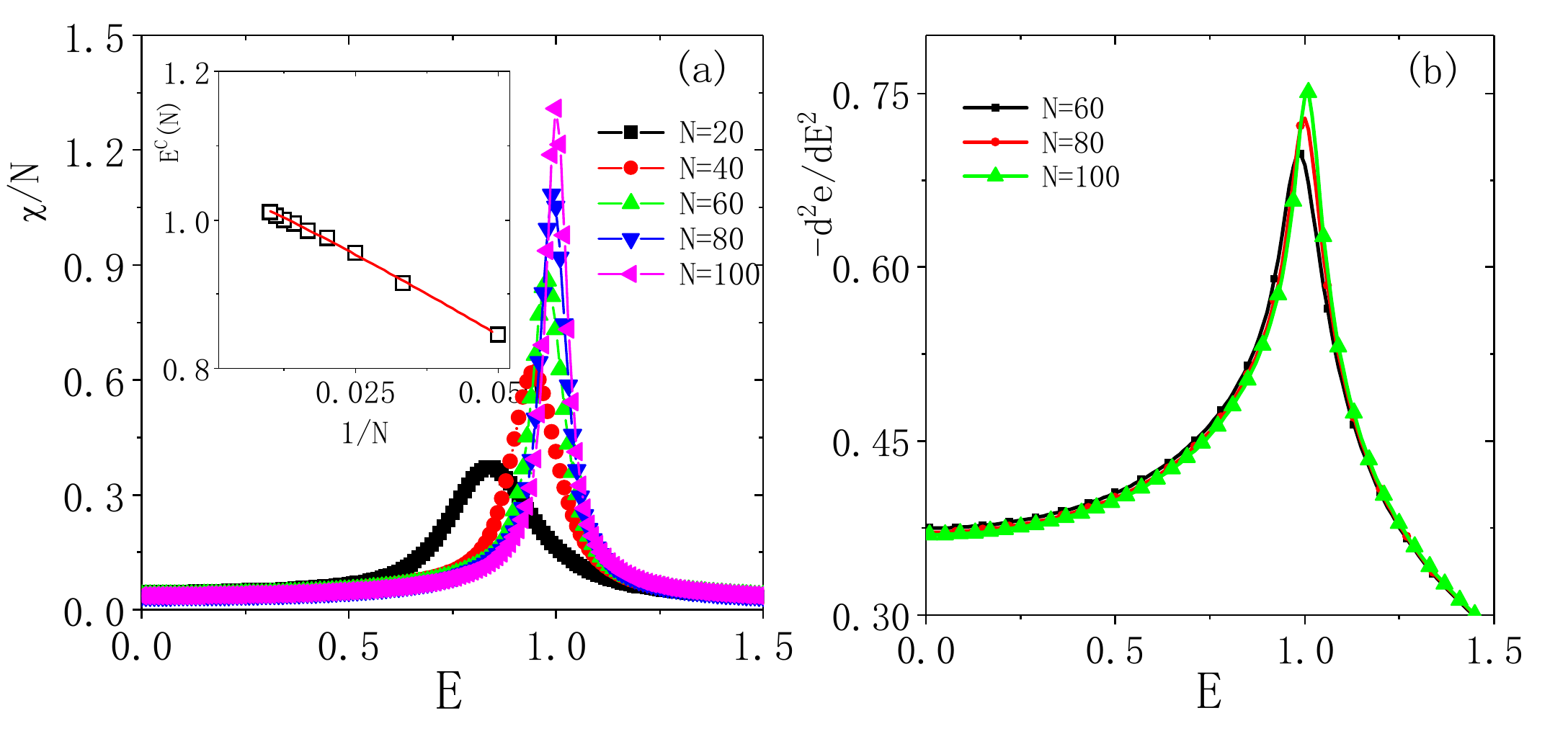}
\caption{(a) Fidelity susceptibility per site and (b) the second-order derivative of the ground-state energy density are plotted as a function of the rhombic single-ion anisotropy $E$ for different system sizes $N$ with $\Delta=1.5$. Inset in (a) shows finite-size scaling behavior of $E_{c}$ in terms of the fidelity susceptibility.
\label{fig8}}
\end{figure*}

\subsection{Case of $\Delta=1.5$}

Furthermore, we also study the IPR of the ground-state with $\Delta=1.5$ in Fig. \ref{fig6}. When the system is in the Z-N\'{e}el phase, the IPR is small and nearly size-independent; see Fig. \ref{fig2_scaling}. 
When the system enters the Large-$E_x$ phase, the IPR grows sharply. The peak in the IPR indicates a QPT between the N\'{e}el phase and the Large-$E_x$ phase. We identify the critical point $E^{c}_3=1.04, a_4=-6.716, b_4=-1.261$ by fitting with Eq.(\ref{eq9}). Besides, the EE is plotted as a function of the rhombic single-ion anisotropy $E$ for different system sizes with $\Delta=1.5$ in Fig. \ref{fig7}(a). One observes that the EE initially increases and then declines. The peak of EE, together with the extreme points identified by the first-order derivation of the local QC in Fig. \ref{fig7}(b), can pinpoint the critical point.

We also calculate the ground-state FS for system size $N$ up to $100$ with $\Delta=1.5$. The ground-state FS per site $\chi/N$ is plotted as a function of the rhombic single-ion anisotropy $E$ for different sizes in Fig. \ref{fig8}(a). The peak's value of $\chi/N$ increases when the system size increases. Similar to the isotropic case, the peak's location also shifts upward as the system size $N$ increases. The scaling of the extreme points of the FS is shown in the inset of Fig. \ref{fig8}(a). We find that the scaling of maximal points can also be fitted by Eq.(\ref{eq9}), which gives $E_{3}^{c}=1.045, a_5=-5.41, b_5=-1.10$. In Fig. \ref{fig8}(b), the second-order derivative of the ground-state energy density implies that the transition between the N\'{e}el phase and large $E_x$ phase is of the second order.

\begin{figure}[t]
\includegraphics[width=0.40\textwidth]{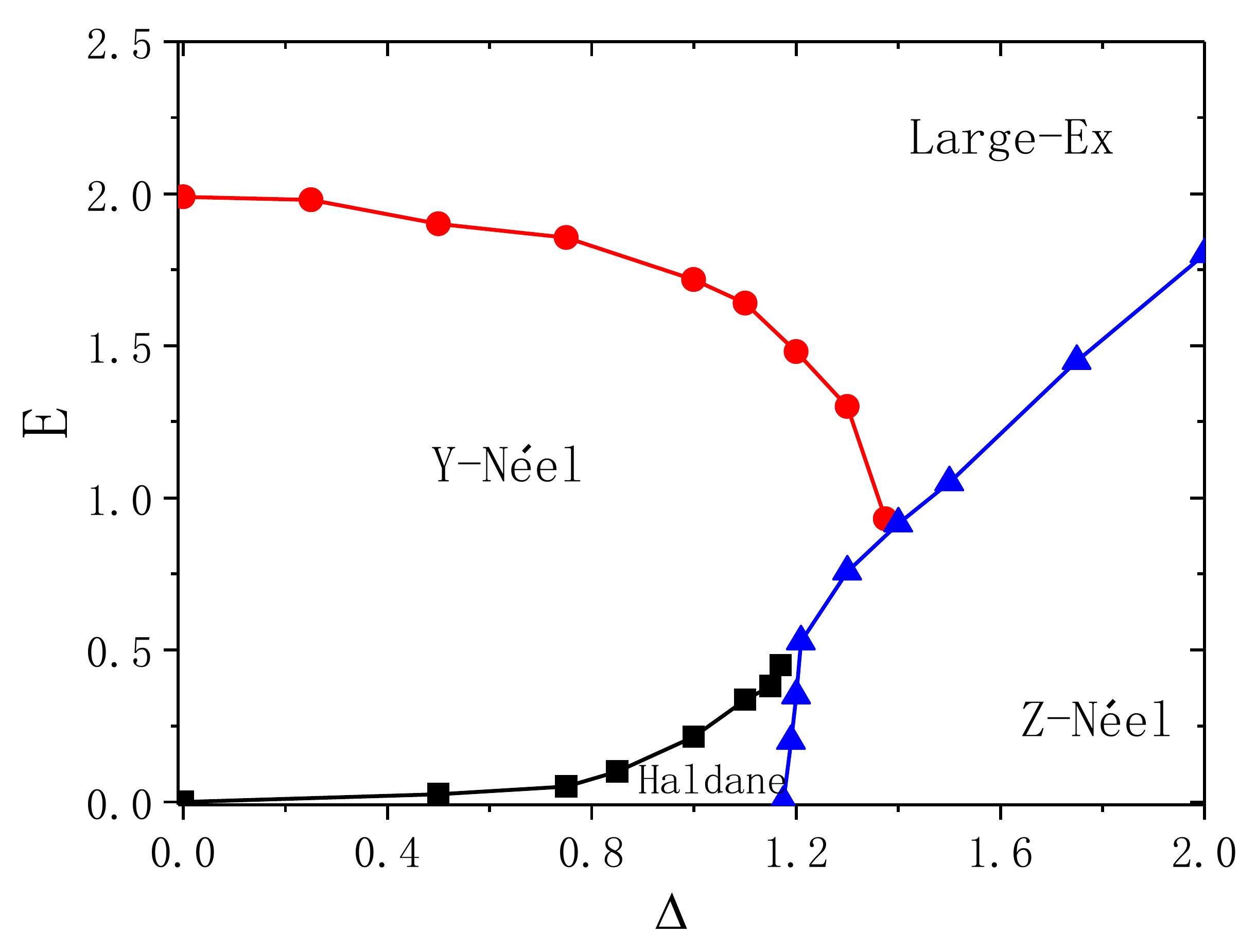}
\caption{ Phase diagram of spin-1 XXZ chain as functions of the rhombic single-ion anisotropy $E$ and the exchange anisotropy $\Delta$.
\label{fig9}}
\end{figure}

We compared the critical values extracted by all the measures in Table \ref{table1}, and we find at least the second digit after the decimal point of those extracted values agree with each other.
To this end, we portray the $E$-$\Delta$ phase diagram of the
Hamiltonian (\ref{eq1}) in terms of the above-mentioned measures. As shown in Fig. \ref{fig9},
the Haldane phase switches to the Z-N\'{e}el phase when $\Delta$ exceeds a critical value $\Delta^c \equiv 1.175$ for $E=0$ \cite{Nomura,Sakai,Ren01}, and $\Delta^c$ obtains a slow increase when $E$ rises from zero. When $\Delta=0$, an infinitesimal rhombic single-ion anisotropy renders the ground state being in the Y-N\'{e}el phase, and the system changes to the Large-$E_x$ phase after $E$ surpasses $E^c\equiv 2$. As $\Delta$ increases, the critical point separating the Y-N\'{e}el phase and the Haldane phase
decreases, unitl it reahes the terminal point of $\Delta^c$=1.37, $E^c=0.93$, which sets a tricritical point.
In contrast, the critical line separating Y-N\'{e}el phase and the Large-$E_x$ phase grows with the increase of
$\Delta$, and it meets the critical line between the Haldane phase and the Z-N\'{e}el phase at $\Delta^c=1.2, E^c=0.53$, which establishes
another tricritical point. We also note that the critical line between the Z-N\'{e}el phase and the large-$E_x$  phase would approach $ E = \Delta$ for large $\Delta$.
\begin{table}
\begin{tabular}{|c|c|c|}
  \hline
 Criterion   &~ $E_1^c~$ & ~$E_2^c$~ \\\hline
 ~$\frac{\partial^2 e}{\partial E^2}$ ~ &~ 0.211 ~ & ~ 1.705~ \\\hline
 ~$T$or$\frac{dT}{dE}$ &~ 0.21~ & ~1.723~ \\\hline
  $S$ or $\frac{dS}{dE}$  & 0.217 & 1.712 \\\hline
  $\chi$ & 0.212 & 1.705 \\  \hline
  QC & 0.21  & 1.72  \\
  \hline
\end{tabular}
\caption{ Comparison of critical points  obtained by different measures referred in the paper for $\Delta=1$.
\label{table1}}
\end{table}

\vspace{0.3cm}
\section{Conclusions }
\label{sec:Discussion}

In this paper, we have numerically investigated the quantum
phase transitions in the one-dimensional
spin-1 XXZ chain with the rhombic single-ion anisotropy by
analyzing a few information
theoretical measures, including the bipartite entanglement entropy, the
fidelity susceptibility and the Wigner-Yanase skew information in addition to other order parameters. Their relation with quantum phase transitions
is discussed. It is important to emphasize that the quantum
phase transitions from the Haldane phase to the Y-N\'{e}el phase, and
the phase transitions from the Y-N\'{e}el phase to the large-$E_x$ phase can be well characterized by the fidelity susceptibility.
The finite-size scalings predicts that the fidelity susceptibility should diverge
in the thermodynamic limit at the pseudo-critical points and
the locations of extreme points approach the real quantum critical
point accordingly. We identify that these quantum phase transitions are of second order by the second-order derivative of the ground-state energy.
Conclusions drawn from various quantum information observables agree well with each other. Finally we provide a ground-state phase diagram as functions of the exchange anisotropy $\Delta$ and the rhombic single-ion anisotropy $E$.
To sum up, the information theoretical measures are effective tools for detecting diverse quantum phase transitions in spin-1 models.
\vspace{0.3cm}
\begin{acknowledgments}

This work is supported by the National Natural Science
Foundation of China (NSFC) under Grants No.~11374043, No.~11474211 and No.~11404407, as well as the Natural Science Foundation
of Jiangsu Province of China under Grant  No.~BK20140072.

\end{acknowledgments}

\end{document}